\documentstyle[11pt,paspconf,epsf]{article}
\newcommand{\mcol}[3]{\multicolumn{#1}{#2}{#3}}

\newcommand{\Msol}{\mbox{\rm M$_{\odot}$}}
\begin{document}
\title{The usage of the Asymptotic Giant Branch Star 
       Features As Age Indicators in Post-Starburst 
       Galaxies}

\author{M. Mouhcine, A. Lan\c{c}on}
\affil{Observatoire Astronomique de Strasbourg.\\ 
11, rue de l'Universit\'e, F-67000 Strasbourg, France}

\begin{abstract}
We investigate techniques that can be used to determine ages of 
recently star forming regions during the phase dominated in the 
near-IR by the asymptotic giant branch stars ($10^8-10^9$ yrs). 
In particular, we present selected near-IR spectroscopic features 
that identify the contribution of O-rich and C-rich AGB 
variable stars to the integrated spectra of post-starbursts. The 
observational strategy based on those features is presented. We 
discuss the robustness of our selected features in constraining 
the ages of the post-starburst population depending on its physical 
environment and on underlying evolved populations. {\it{The 
interplay between the integrated features of populations and the 
stellar parameters is discussed.}}
\end{abstract}


\section{Introduction}
It is well established that the asymptotic giant branch stars (AGB stars) 
contribute a large fraction of the integrated near-IR light of stellar 
populations at some stages of their evolution (Persson et al 1983, Bruzual 
\& Charlot 1993).  While methods have been suggested to investigate the age 
of younger population dominated by evolved massive stars 
(Mayya 1997), there has been until recently no technique in the literature 
to recognize the AGB contribution in the integrated spectra of stellar 
populations (Lan\c{c}on et al. 1999). This is 
complicated by the fact that red giant stars and AGB stars display 
similar CO bands in their near-IR spectra (e.g. at 2.3 $\mu$m), and it is
very difficult to distinguish between both populations using only such 
features. Since two decades there have been many observations about the AGB in 
the galaxies of the Local Group (Cook et al. 1986, Frogel et al. 
1990) but there is as yet no systematic investigation to 
detect AGB stars in galaxies beyond the local group, a task that could only
rely on integrated emission properties. The scope of this
work is to introduce in accurate way the relevant particularities of 
evolution through the AGB phase, taking into account a ``detail" which 
was completely ignored until now and which strongly affects the spectroscopic 
signatures of the AGB stars: the variability.

\section{Modelling the spectral evolution}
In our calculations the stars evolve up to the end of the early AGB
following the tracks calculated by Bressan et al (1993) and 
Fagotto et al (1994). The AGB evolution beyond this stage is characterized 
by two types of pulsations: 
(i) the secular thermal pulses and (ii) the Mira-like 
pulsation with a much shorter time scale (P $\sim 10^2 - 10^3$ days). 
To study the AGB contribution to the integrated spectrophotometric properties,
one must include the effects of the two phenomena in the population
synthesis models.
An interesting attempt to introduce the first effect was proposed by 
Girardi \& Bertelli (1998), but they did not include the chemical 
evolution of the AGB stars, and the effect of the envelope burning was 
parametrized in an approximative way. To go one step further we have 
constructed a synthetic model for AGB evolution that includes
the relevant processes (mass loss, envelope burning,
dredge-up). The second type of pulsation has a strong effect on the spectra of 
AGB stars. In pre-existing population synthesis models, the separation 
between static giants and AGB stars was ignored in the input stellar libraries. 
Since the atmosphere models for Miras are still far from being satisfactory 
(Mouhcine \& Lan\c{c}on 1998), we have decided to introduce the short time 
scale pulsation by the use of a new empirical stellar library 
constructed from a large sample of multiple instantanous near-IR spectra 
(wavelength coverage from 0.5 to 2.5 $\mu$m), that contains 
O-rich and C-rich stars. For more information on the construction of this
library see Mouhcine \& Lan\c{c}on (this conference) and 
Mouhcine \& Lan\c{c}on 1999 (in preparation). 

\section{Strategy to look for AGB stars in post-starburst systems}
The observational strategy that one may adopt to search for spectroscopic 
signatures of AGB stars should reveal them whatever their surface chemistry,
but could be based on the spectral differences between near-IR spectra of 
C-rich and O-rich stars for finer age estimates. 
The O-rich AGB variables display deeper
near-IR H$_{2}$O absorption bands (1.7-2.1\,$\mu$m) than the static red
stars, even at relatively high effective temperatures.
The near-IR spectra of carbon 
stars are dominated by bands of C$_{2}$ and CN, whereas 
those of oxygen-rich stars are dominated by metal oxide bands 
(TiO, VO and H$_{2}$O). \\
Among commonly used indices, our calculations show that the H$_{2}$O index 
at $2\mu$m is the most sensitive one and shows a large variation with time; 
however this index is unable to detect carbon stars, which can emit a 
large part of the bolometric luminosity of the AGB star population since they 
are in general at the tip of the asymptotic giant branch, particularly 
for the metal-poor systems. To overcome this problem 
we include an other spectroscopic index around 1.77 $\mu$m. This index  
identifies both the H$_{2}$O molecular absorption in the 
O-rich AGB stars, and the C$_{2}$ bandhead of C-rich AGB stars.\\ 
Figure 1 shows our favorite diagnostic diagram (Lan\c{c}on et al. 1999). 
The TP-AGB lifetime fraction spent as a C-rich object, as a function
of initial mass, is taken from Groenewegen \& de Jong (1993, 1994), 
who assume a Reimers law to compute the mass loss rate. 
We see clearly in this diagram that the (H$_{2}$O+C$_{2}$) 
index signals the predominance of the AGB stars in the post-starburst 
phase, without any confusion with older or younger populations.   
\begin{figure}
\plotfiddle{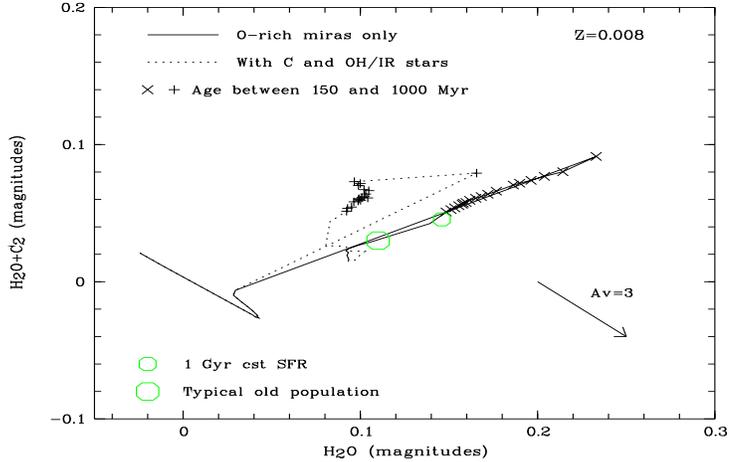}{4.5cm}{-90}{45}{35}{-190}{200}
\caption{Evolution in the prefered molecular index diagram.Cercles 
         identify the predicted location of contaminating older populations.} 
\end{figure}

\section{Discussion }
Many uncertain physical processes control AGB evolution
(mass loss, dredge-up, mixing, hydrogen burning at the  base of the 
convective envelope ... etc). 
There are many theoretical evidences that the long 
timescale evolution along the AGB depends strongly on the way the 
interplay between those processes is introduced in the modelling (Frost 
et al. 1998). In addition, the temperature scales of oxygen-rich and 
carbon-rich TP-AGB stars are poorly known. Uncertainties in dating 
estimates come both from the the individual processes and the interplay 
between them. 
To look into the effect of the stellar input on the temporal 
evolution of molecular features, we test the effect of the mass loss 
formulation and of the temperature scale used to connect the stellar 
spectra to the H-R diagram. Mass loss plays a dominant role in determining 
the lifetime and the extent of nuclear processing on the AGB. 
We have assumed different mass loss prescriptions and computed the 
lifetime and luminosity along the TP-AGB. Table 1 illustrates how much
the predictions may differ. 
The Vassiliadis \& Wood (1993) and Baud \& Habing (1983) prescriptions 
reduce the lifetime drastically in comparison with those estimated using the 
formulation of Bl\"{o}cker \& Sch\"{o}nberner (although the latter $\dot{M}$
is a steep function of the luminosity).
\begin{figure}
\plotfiddle{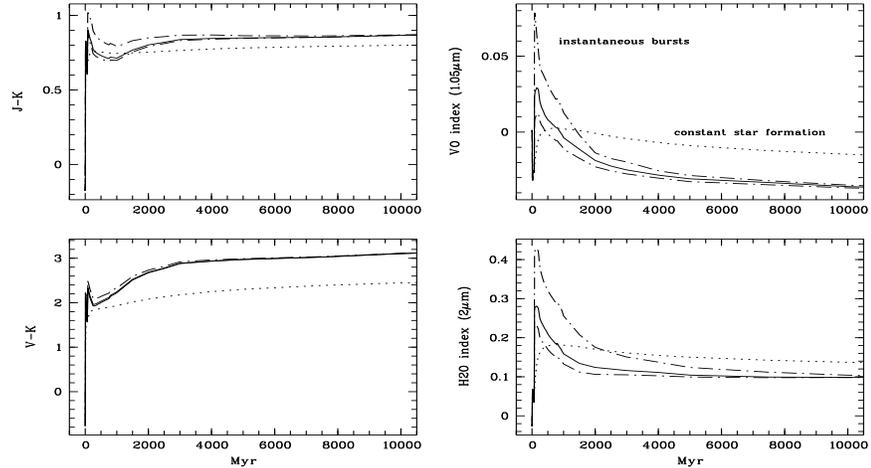}{4.5cm}{-90}{45}{35}{-190}{200}
\caption{Evolution of selected properties at Z=Z$_{\odot}$. 
The lower dot-dashed lines are calculated using Feast's temperature scale, 
the upper ones using the static star temperature calibration (see text). 
} 
\end{figure}
The temperature scale adopted controls the wavelength dependant 
contribution of the AGB population and the evolution of the molecular 
signatures originating from the AGB stars. To test this, we have adopted two 
different scales: the first is suggested by Feast (1996) for 
Mira-type variables, the second is obtained from a static giant model 
atmosphere grid (Alvarez et al, in preparation). Figure 2 shows the effect on 
the temporal behavior of some spectrophotometric quantities. It is clear that 
the molecular indices are better age indicators than broadband colours and
are also more sensitive to the adopted temperatures. 
The higher the temperature assigned to each 
spectrum of our library, the stronger the resulting integrated molecular 
signatures. This offers us an interesting opportunity to constrain 
\pagebreak

the 
temperature scale of AGB stars, using extra-galactic star clusters with well 
constrained ages (such observations are on the way by our group).    

%%%%%%%%%%%%%%%
\begin{table}
\caption{AGB life duration normalized to life duration calculated 
assuming Reimers law. Mass loss prescriptions:
Vassiliadis \& Wood (1993: V\&W),
Bl\"{o}cker \& Sch\"{o}nberner (1995: B\&S),
Baud \& Habing (1983: B\&H).}
\begin{center}
\scriptsize
\begin{tabular}{crrr||rrr}
&\mcol{3}{c}{Z=Z$_{\odot}$} & \mcol{3}{c}{Z=Z$_{LMC}$} \\ 
\tableline
M($\Msol$) &$V\&W$ &$ B\&S $ &$B\&H$ &$V\&W$& $ B\&S $ & $ B\&H$ \\
\tableline
2          & 1.3   & 4.6     & 1.8   & 1.8  & 3.4      &  1.6    \\
2.5        & 1.77  & 3.8     & 1.77  & 2.1  & 3.5      &  1.7    \\
3          & 1.9   & 4.4     & 1.9   & 2.4  & 4.3      &  1.8    \\ 
\end{tabular}
\end{center}

\end{table}
%%%%%%%%%%%%%%%

\end{document}